\documentstyle[11pt,newpasp,twoside,epsf]{article}
\def\edcomment#1{\iffalse\marginpar{\raggedright\sl#1\/}\else\relax\fi}
\reversemarginpar

\begin{document}
\title{Extreme BAL Quasars from the Sloan Digital Sky Survey}
\author{Patrick~B.~Hall,~J.~E.~Gunn,~G.~R.~Knapp,~V.~K.~Narayanan,~M.~A.~Strauss}
\affil{Princeton University Observatory, Princeton NJ 08544-1001}
\author{S. F. Anderson}
\affil{University of Washington}
\author{D. E. Vanden Berk}
\affil{Fermi National Accelerator Laboratory}
\author{T. M. Heckman, J. H. Krolik, Z. I. Tsvetanov, W. Zheng}
\affil{The Johns Hopkins University}
\author{G. T. Richards, D. P. Schneider}
\affil{The Pennsylvania State University}
\author{X. Fan}
\affil{Institute for Advanced Study}
\author{D. G. York}
\affil{The University of Chicago}
\author{T. R. Geballe}
\affil{Gemini Observatory}
\author{M. Davis}
\affil{University of California at Berkeley}
\author{R. H. Becker}
\affil{Lawrence Livermore National Laboratory}
\author{R. J. Brunner}
\affil{California Institute of Technology}
\begin{abstract}
The Sloan Digital Sky Survey 
has discovered a population of 
broad absorption line 
quasars with various extreme properties.  
Many show absorption from metastable states of Fe\,{\sc ii} with varying
excitations; 
several objects are almost completely absorbed bluewards of Mg\,{\sc ii};
at least one shows stronger absorption from Fe\,{\sc iii} than Fe\,{\sc ii},
indicating temperatures $T$$>$35000~K in the absorbing region;
and one object even seems to have broad H$\beta$ absorption.
Many of these extreme BALs are also heavily reddened, though `normal' BALs
(particularly LoBALs) from SDSS also show evidence for internal reddening.
\end{abstract}

\section{Introduction}

The Sloan Digital Sky Survey (York et al. 2000) is using dedicated instruments
on a 2.5m telescope (Gunn et al. 1998) to image 10$^4$ deg$^2$ of sky to
$\sim$23$^m$ in five bands (Fukugita et al. 1996)
and obtain spectra of $\sim$10$^6$ galaxies and $\sim$10$^5$ quasars
selected primarily as outliers from the stellar locus.
Its area, depth, and selection criteria make SDSS effective at finding
unusual quasars. 
The first data release (Stoughton et al. 2001, in prep.). contains
$\sim$4500 spectroscopically confirmed quasars, including $\sim$200 BALs,
a few percent of which have extreme properties of one sort or another.
All these extreme BALs are LoBALs, which show absorption from both low-
and high-ionization transitions, instead of the more common HiBALs with
only high-ionization absorption.  Full analysis is underway 
(Hall et al. 2001, in prep.), but already these objects confirm 
the existence of a population of extreme BALs,
as suspected from previous discoveries of individual extreme BALs
(Becker et al. 1997, Djorgovski et al. 2001).

\section{BAL Quasars With Fe\,{\sc ii}* Absorption}

\begin{figure}
\plottwo{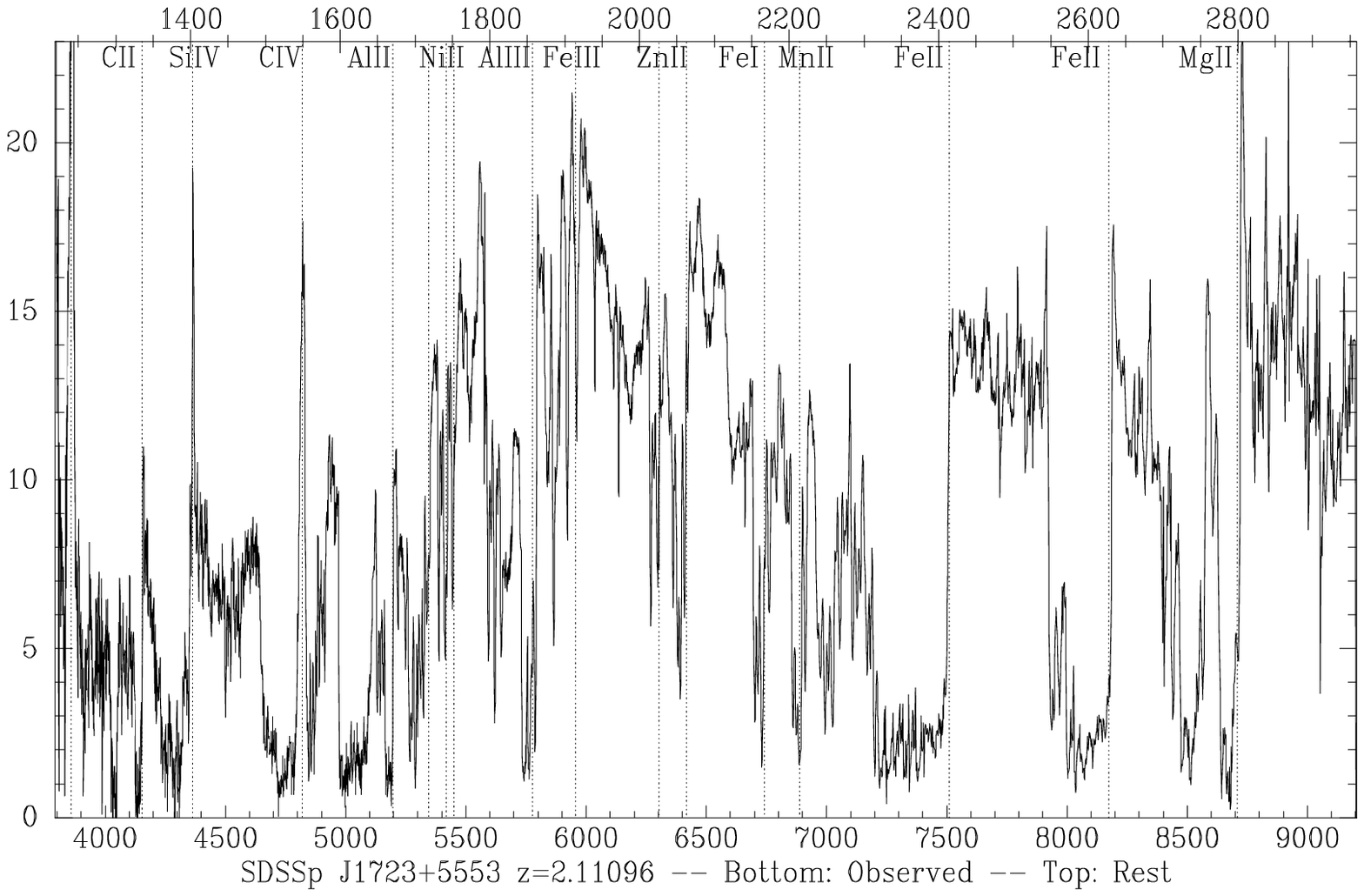}{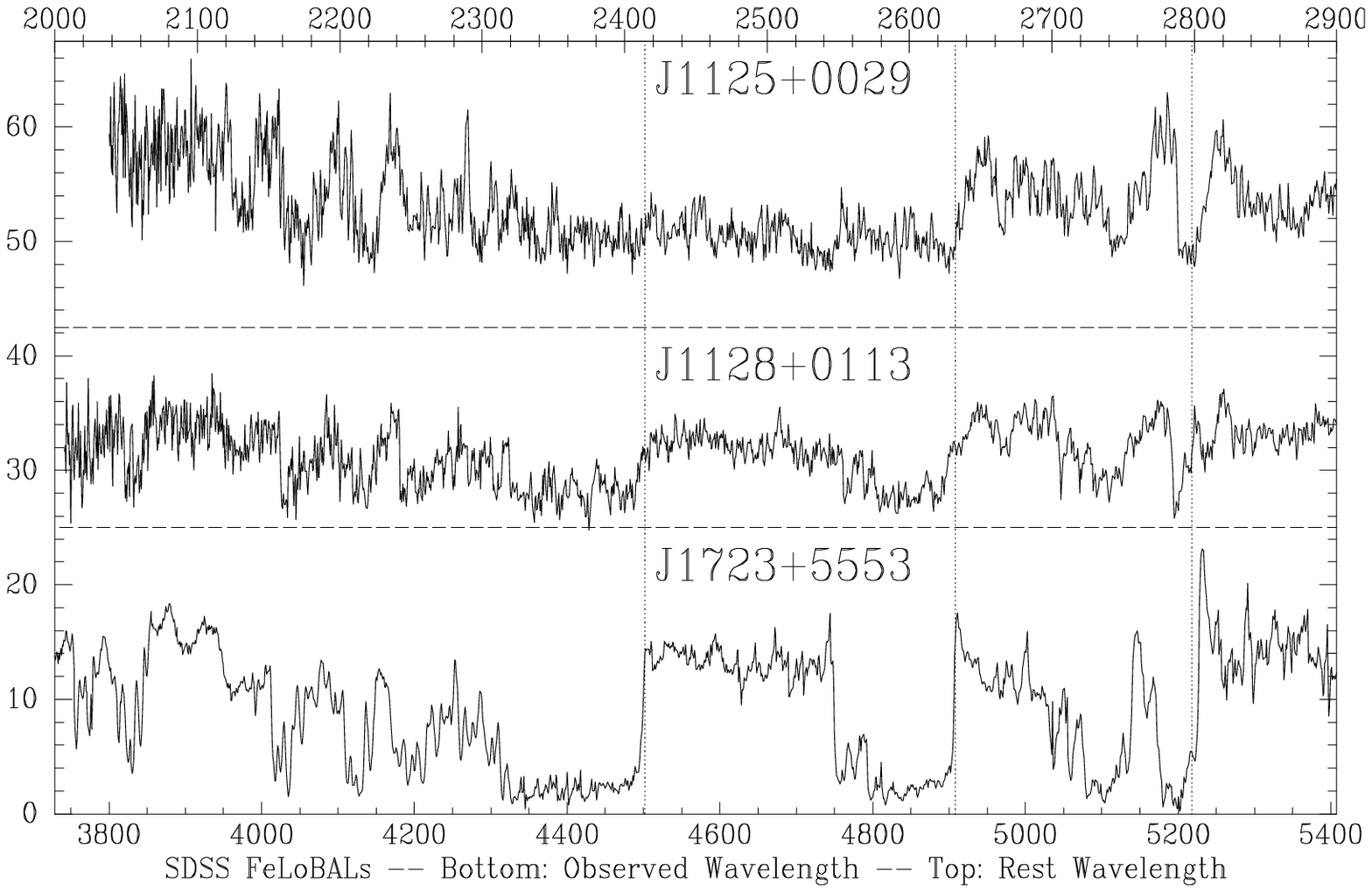}
\caption{BALs with Fe\,{\sc ii}* absorption. a) SDSS 1723+5553, $z$=2.11.
b) Comparison of 2000-2900\AA\ regions in SDSS~1723+5553 (bottom)
and two lower-$z$ FeLoBALs, SDSS~1128+0113 (middle, $z$=0.894)
and SDSS~1125+0029 (top, $z$=0.865); dashed lines show zero flux for each,
and vertical dotted lines show 
Fe\,{\sc ii}*2414\AA\ \& 2632\AA\ and Mg\,{\sc ii} 2798.}
\end{figure}

The rare 
LoBAL quasars with absorption from metastable excited states of
Fe\,{\sc ii} (Fe\,{\sc ii}*) 
	have been dubbed FeLoBALs 
	(Becker et al. 2000; Hazard et al. 1987; Menou et al. 2001).
	They
are valuable because photoionization modelling of them can 
constrain $n_e$ 
in the BAL clouds (e.g. de Kool et al. 2001).
%
Fig. 1a shows a spectacular example, SDSS~1723+5553,
with absorption from over twenty transitions in at least a dozen elements.
Fig. 1b compares SDSS~1723+5553 to two lower-$z$ FeLoBALs with [O\,{\sc ii}]
emission line redshifts.  
Both low-$z$ objects show Fe\,{\sc ii}* absorption blueward of
2414\AA\ and 2632\AA\ from 
states up to $\sim$1~eV
above ground, but SDSS~1125+0029 
also shows 
absorption near 2500~\AA\ 
from even more excited levels.  
In both low-$z$ objects, the Mg\,{\sc ii} BAL absorption apparently
extends 2000~km\,s$^{-1}$ {\em redward} of the 
systemic $z$. 


\begin{figure}
\plottwo{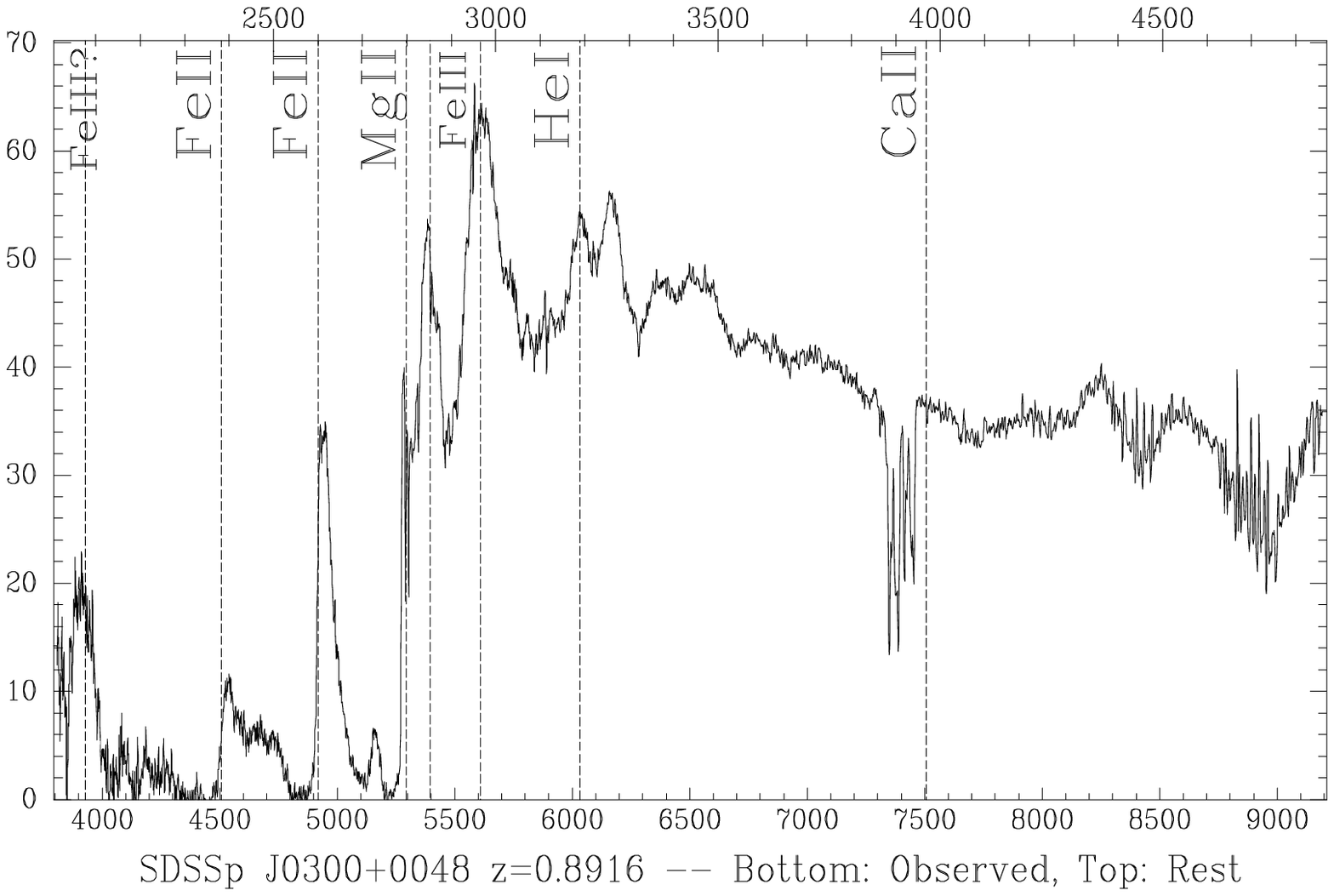}{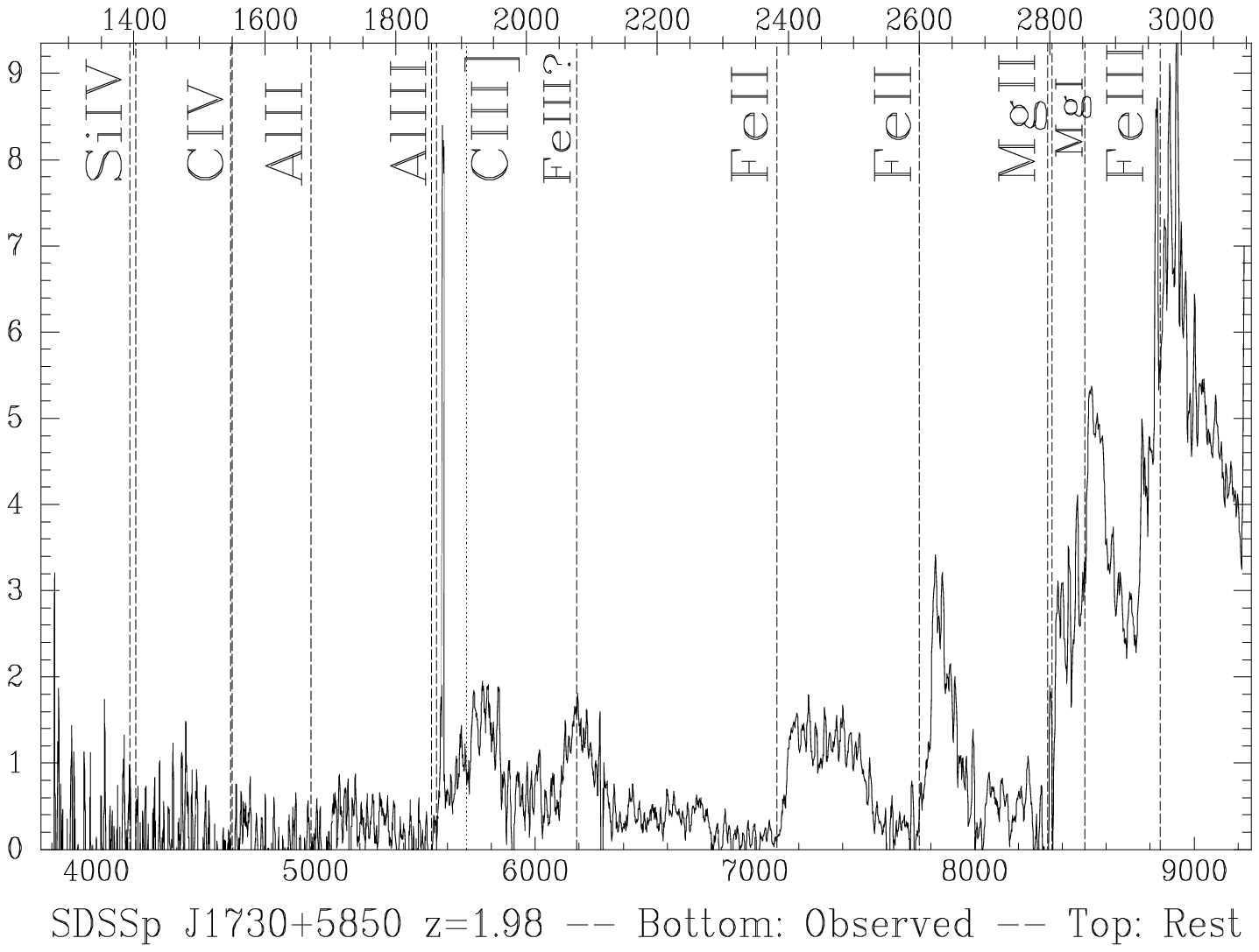}
\caption{SDSS LoBALs with strong absorption blueward of Mg\,{\sc ii}.
a) SDSS~0300+0048 at $z$=0.8916.
b) SDSS~1730+5850 at $z$=1.98.}
\end{figure}

Some SDSS FeLoBALs show very abrupt drops in flux near
Mg\,{\sc ii}\,$\lambda\lambda$2796,2803 (e.g. SDSS~0300+0048 in Fig. 2a).
%
SDSS~0300+0048 has associated Mg\,{\sc ii} absorption at $z$=0.8916, 
at least 4 narrow Ca\,{\sc ii} H\&K absorption systems located 2350 to
3900~km\,s$^{-1}$ blueward of the Mg\,{\sc ii} system, and
broad Ca\,{\sc ii} absorption extending a further 2000~km\,s$^{-1}$ blueward.
Broad, near-total Mg\,{\sc ii} absorption is associated with the highest-$z$
Ca\,{\sc ii} system, but broad Fe\,{\sc ii}* absorption 
is associated instead with the {\em strongest} Ca\,{\sc ii} system,
at slightly lower $z$.
Fig. 2b shows SDSS~1730+5850, which is clearly a higher redshift analogue of
SDSS~0300+0048.  Our spectrum extends farther into the UV for this $z$$\sim$2
object, and shows a weak recovery at C\,{\sc iii}]\,$\lambda$1908 but
essentially zero flux below Al\,{\sc ii}\,$\lambda$1670.
%
Quasars such as these at $z\geq2$ will obviously be greatly
underrepresented in optical surveys. 

\section{BAL Quasars With Fe\,{\sc iii} Absorption}

\begin{figure}
\plotone{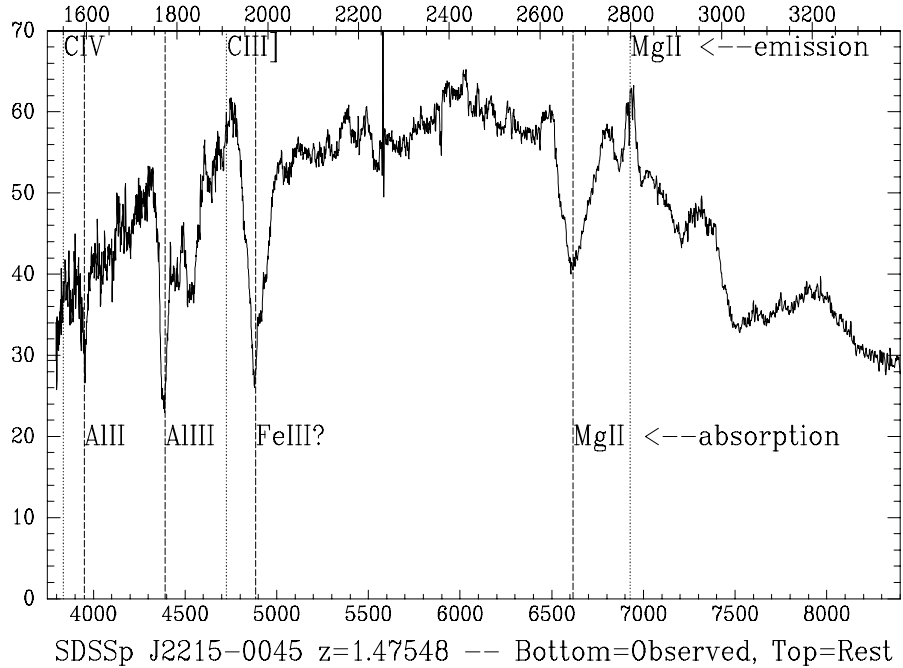}
\caption{SDSS~2215$-$0045 at $z$=1.47548.  Note the absence of strong
Fe\,{\sc ii} absorption at 2200--2600\AA\ in the rest frame.
}
\end{figure}

Fig. 3 shows SDSS~2215$-$0045, a 
LoBAL at $z$=1.47548
(measured from associated Mg\,{\sc ii} absorption, as with SDSS~0300+0048).
Its absorption troughs are unusual for a LoBAL:
they are very broad, 
detached, and strongest near the high velocity end rather than at low velocity.
By comparison to SDSS~1723+5553 (Fig. 1), 
we initially identified the strong
trough at $\lambda_{obs}$$\sim$4900\AA\ as Cr\,{\sc ii}.  However, the implied
abundance of Cr relative to Mg is implausible, and 
the expected corresponding Zn\,{\sc ii} is missing.
We now believe this absorption is due to Fe\,{\sc iii} (multiplet UV 48), with
additional Fe\,{\sc iii} (UV 34) absorption at $\lambda_{obs}$$\sim$4500\AA,
redward of Al\,{\sc iii}.
Since Fe\,{\sc ii} absorption is weak or absent,
the large Fe\,{\sc iii}/Fe\,{\sc ii} ratio suggests that the BAL clouds
in this object have $T$$>$35000~K, sufficient to collisionally ionize 
Fe\,{\sc ii} to Fe\,{\sc iii}.
Fe\,{\sc iii} absorption is seen in several other SDSS LoBALs (e.g. Fig. 4)
and in a few previously known LoBALs,
but nowhere as strongly (alone or relative to Fe\,{\sc ii})
as in SDSS~2215$-$0045.
Note the different spectral slopes blueward \& redward of $\sim$2400\AA,
indicating reddening which must occur outside the BAL region since dust
cannot survive long at $T$$>$35000~K. 



\section{A LoBAL With Broad H$\beta$ Absorption}

\begin{figure}
\plotone{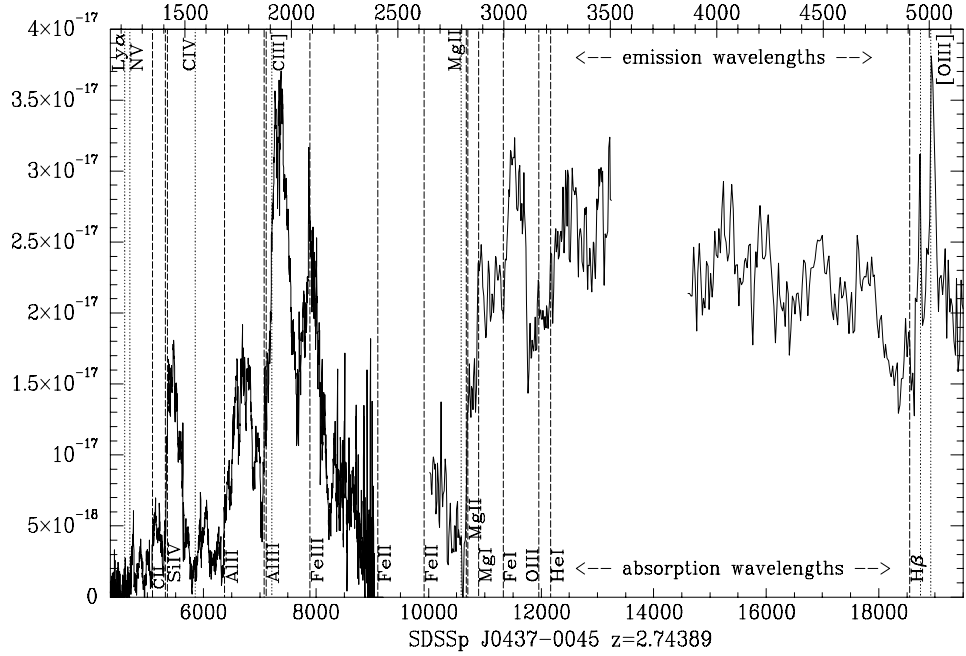}
\caption{Optical-NIR spectra of SDSS~0437$-$0045 at $z$=2.74389.
Strong lines are labelled at the expected wavelengths for emission (top)
and absorption (bottom).  Note the nearly complete absorption near the expected
wavelength of C\,{\sc iv}, and the probable broad H$\beta$ absorption.}
\end{figure}

Fig. 4 shows an optical (Keck) plus NIR (UKIRT) spectrum of SDSS~0437-0045
which reveals a strongly absorbed quasar with $z$=2.74389 from [O\,{\sc iii}].  
The absorption extends 2900~km\,s$^{-1}$ redward of this $z$ (cf. Fig.~2b).
Even more remarkable is the probable presence of H$\beta$ absorption
nearly 10$^4$~km\,s$^{-1}$ wide and of REW$\sim$100~\AA.
H$\beta$ absorption in AGN has previously been seen only in NGC~4151
(Anderson \& Kraft 1969; Sergeev et al. 1999),
but with 
$\leq$1000~km\,s$^{-1}$ width and $\leq$3~\AA\ REW.
This object is also unusual because the Fe{\sc iii} 
trough at $\sim$2070\AA\ has been seen to vary with nearly unprecendented
amplitude and speed.

\section{Heavily Reddened BAL Quasars}

\begin{figure}
\plotone{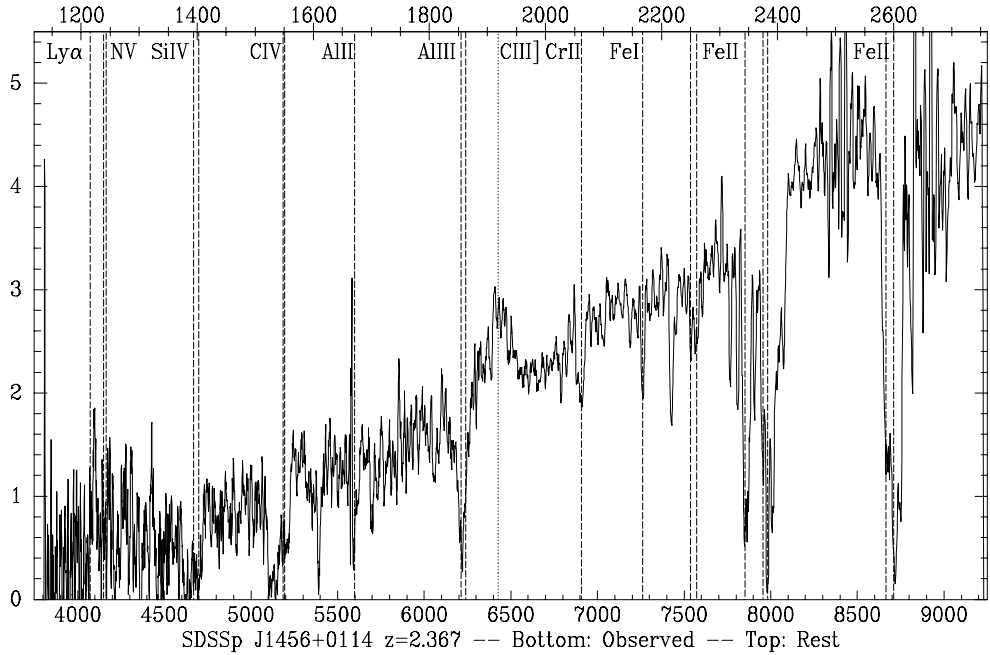}
\caption{SDSS~1456$+$0114, an extremely reddened LoBAL at $z$=2.367.}
\end{figure}

SDSS has found evidence for a population of red quasars (Richards et al. 2001),
and BALs in general are redder than the typical quasar (Menou et al. 2001),
but for the most extreme objects the reddening is unambiguous.
Fig. 5 shows SDSS~1456+0114, an extremely reddened LoBAL 
(and FIRST source) at $z$=2.367  
(measured from weak C\,{\sc iii}], the only broad line visible).
Several other similar objects have been found by SDSS, but with even
weaker broad emission.  Since reddening does not affect equivalent widths,
this may indicate that in these objects the broad line region is even more
heavily reddened than the continuum.  This would be quite plausible if most
of the continuum light in those objects is scattered light, a hypothesis which
can easily be tested with polarization data.

\section{Discussion}

`Well, that was disturbing.' ---Fred Hamann, after this talk at the meeting.

We prefer to view these objects as invigorating.  
The area and depth of SDSS, plus its simple selection of quasar candidates as
outliers from the stellar locus, makes it efficient at finding quasars with
unusual properties.  Moreover, the discovery of typically several examples of
each type of extreme LoBAL quasar presented here means that a {\em population}
of extreme LoBAL quasars exists and that only now, with SDSS, are we beginning
to sample the full range of properties that exist in BAL outflows, and
thus around quasars on the whole.
%

\acknowledgements
The Sloan Digital Sky Survey 
	(SDSS)
is a joint project of The University of
Chicago, Fermilab, the Institute for Advanced Study, the Japan Participation
Group, The Johns Hopkins University, the Max-Planck-Institute for Astronomy 
	(MPIA), 
the Max-Planck-Institute for Astrophysics 
	(MPA), 
New Mexico State University, Princeton University,
the United States Naval Observatory, and the University of Washington.
Apache Point Observatory, site of the SDSS telescopes,
is operated by the Astrophysical Research Consortium 
	(ARC).
Funding for 
	the project 
has been provided by the Alfred P. Sloan Foundation,
the SDSS member institutions, the National Aeronautics and Space Administration,
the National Science Foundation, the U.S. Department of Energy, 
	the Japanese
Monbukagakusho, and the Max Planck Society.
The SDSS Web site is http://www.sdss.org/.

 
\end{document}